\documentclass{webofc}
\usepackage[varg]{txfonts}   % Web of Conferences font
\usepackage{xcolor}
\begin{document}
\title{Direct, Resonant Production of States with Positive Charge Conjugation in Electron-Positron Annihilation}
%
% subtitle is optionnal
%
%%%\subtitle{Do you have a subtitle?\\ If so, write it here}

\author{\firstname{Henryk} \lastname{Czy\.{z}}\inst{1} \thanks{Work supported in part by the Polish National Science Centre, grant
number  DEC-2012/07/B/ST2/03867  and  German  Research  Foundation DFG under Contract No. Collaborative Research Center CRC-1044.} \and
        \firstname{Johann} \lastname{H. K\"{u}hn}\inst{2} \and
        \firstname{Szymon} \lastname{Tracz}\inst{1}
        % etc.
}
\institute{Institute of Physics, University of Silesia, PL-40007 Katowice, Poland 
\and
           Institut für Theoretische Teilchenphysik, Karlsruhe Institute of Technology, D-76128 Karlsruhe, Germany 
          }

\abstract{%
 Recent theory results on direct production of resonances with positive charge conjugation in electron-positron annihilation are reviewed. The strong model dependence is emphasized, with predictions varying between 0.03~eV and 0.43~eV for the charmonium state with $J^{PC} = 1^{++}$ and between 0.16~eV and 4.25~eV for the state with $J^{PC} = 2^{++}$. For the state with $J^{PC} = 0^{++}$ the cross section is of $\cal{O}$ $(m^2_e/M^2_{\chi})$ and thus negligeable
for all practical purpose. The importance of the relative phase of the
production amplitude is emphasized.
}
\maketitle
\section{Introduction}
\label{intro}
Resonant production of quarkonium states with the quantum numbers
$J^{PC}=1^{--}$ has been of interest for experiments at electron-positron colliders from the beginning. Great emphasis has been put on charmonium and bottomonium ground- and excited states. Their  production rates are
proportional to the widths of these states into electron-positron pairs and thus --- in the framework of the nonrelativistic quark model ---
proportional to the square of the wave function at the origin. In principle
also states with $J^{PC} = 1^{++}$ and $2^{++}$ can be produced in
electron-positron annihilation. However, in this case the production proceeds
through two virtual photons. Compared to the $S$-waves with
$J^{PC} = 1^{--}$ this leads, obviously, to a suppression of the production rate by a factor $\alpha^2$. In the short distance approximation the coupling is, furthermore, proportional to the derivative of the wave function at the origin, which implies a further suppression by a factor $(v^2/c^2) \approx {\cal O}(0.1)$. In total one thus expects a reduction of the enhancement by a factor between ${\cal O}(10^{-5})$ and ${\cal O} (10^{-4})$.

The first analysis of this process \cite{Kuhn:1979bb,Kaplan:1978wu} was based on the short-distance approximation only.
In the meantime various modifications and improvements have been
formulated \cite{Yang:2012gk,Kivel:2015iea,Denig:2014fha,Czyz:2016xvc}. These papers, however, have extended the spread of the predictions considerably. Two production
channels are of particular interest: The process
$e^+e^- \to \chi_J \to hadrons$
and the process
$e^+e^- \to \chi_J \to J/\psi + \gamma$ with the subsequent decay of $J/\psi$ into $\mu^+\mu^-$ or $e^+e^-$.
Note, that the interference with the continuum cross section
$e^+e^- \to J/\psi + \gamma$ and thus the relative phase of the process will play an important role in this connection.

In principle there is also the production of the $1^{++}$ state through the neutral current which must be considered. In practice, however, this induces a small additional contribution to the production amplitude only, which barely
affects the cross section.

It is the purpose of this note to recall the basic aspects relevant for this reaction. We will first discuss the amplitude relevant for resonant $\chi_J$ production, employing various approximations for the
$\chi_J\gamma^*\gamma^*$ coupling. In the next step we will consider the full process $e^+e^-\to\gamma\mu^+\mu^-$ with the photon radiated from the initial $e^+e^-$ state (Fig.1a), from the $\chi_J$ state through its decay into $J/\psi$ and a photon (Fig.1b) and from the off-resonant amplitude
(Fig.1c).

After presenting the results of the analytical calculations for the amplitudes we will give the predictions of a Monte Carlo generator for the $\mu^+\mu^-\gamma$ final state, taking the resonant signal, the flat background and the interference into account. 
\begin{figure}[h]
% Use the relevant command for your figure-insertion program
% to insert the figure file.
\centering
\includegraphics[width=8cm,clip]{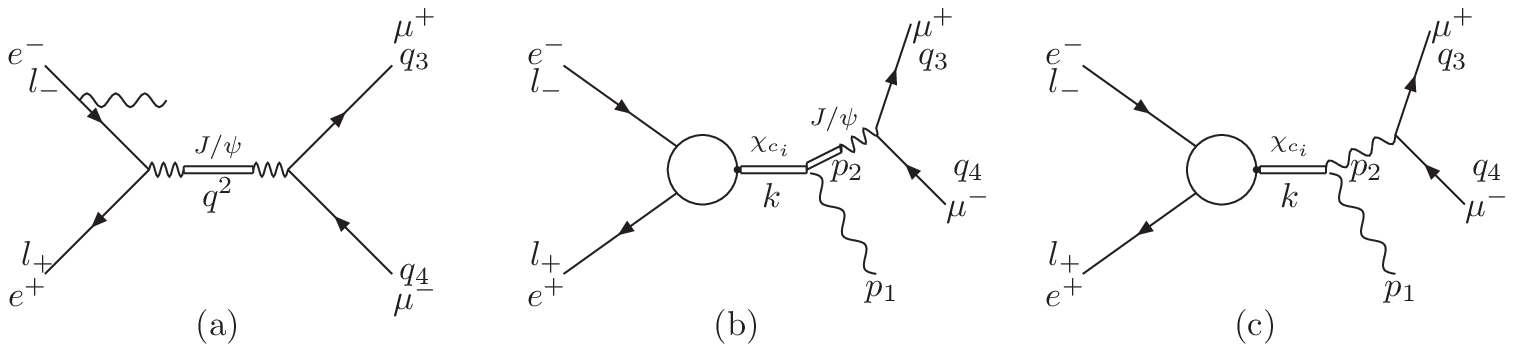}
\caption{Diagrams for the cross section for the process $e^+ e^- \to \chi_{_J}\to \gamma J\psi(\to \mu^+ \mu^-)$.}
\label{fig-1}       % Give a unique label
\end{figure}
\section{Resonant $\chi_J$ Production}
\label{sec-1}
Using the short distance approximation as discussed in \cite{Kuhn:1979bb}, the coupling of $\chi_J$ to two virtual photons is given by
\begin{eqnarray}
\label{eq1}
A_0^{\alpha\beta}(p_1,p_2)\epsilon^1_{\alpha}\epsilon^2_{\beta}&=&\sqrt{\frac{1}{6}}c\frac{2}{M_{\chi_{0}}}\{[(\epsilon_1\epsilon_2)(p_1p_2) \nonumber-(\epsilon_1p_2)(\epsilon_2p_1)][M_{\chi_{0}}^2+(p_1p_2)] \nonumber \\&&+(\epsilon_1p_2)(\epsilon_2p_2)p_1^2 +(\epsilon_1p_1)(\epsilon_2p_1)p_2^2\nonumber-(\epsilon_1\epsilon_2)p_1^2p_2^2 -(\epsilon_1p_1)(\epsilon_2p_2)p_1p_2\nonumber \},\\ 
\end{eqnarray}

\begin{eqnarray}
 \label{eq2}
A_1^{\alpha\beta}(p_1,p_2,\epsilon)\epsilon^1_{\alpha}\epsilon^2_{\beta}&=&ic\{ p_1^2(\epsilon,\epsilon_1,\epsilon_2,p_2)+p_2^2(\epsilon,\epsilon_2,\epsilon_1,p_1)+\epsilon_1p_1(\epsilon,\epsilon_2,p_1,p_2)+\epsilon_2p_2(\epsilon,\epsilon_1,p_2,p_1)\}\nonumber,\\
\end{eqnarray}

\begin{eqnarray}
\label{eq3}
A_2^{\alpha\beta}(p_1,p_2,\epsilon)\epsilon^1_{\alpha}\epsilon^2_{\beta}&=&\sqrt{2}cM_{\chi_{2}}\{(p_1p_2)\epsilon^1_{\mu} \epsilon^2_{\nu}+p_{1\mu}p_{2\nu}(\epsilon_1\epsilon_2)-p_{1\mu}\epsilon_{\nu}^2(\epsilon_1p_2))-p_{2\mu}\epsilon_{\nu}^1(\epsilon_2p_1))\}\epsilon^{\mu\nu}\nonumber, \\
\end{eqnarray}

where 
\begin{eqnarray}
\label{eq4}
c &\equiv&c((p_1+p_2)^2,p_1^2,p_2^2,m)=\frac{16\pi\alpha a}{\sqrt{m}}\frac{1}{((p_1-p_2)^2/4-m^2+i\epsilon)^2}.
\end{eqnarray}

Here $m$ stands for the effective mass of the charmed quark and
$a \equiv \sqrt{\frac{1}{4\pi}}3 Q^2\Phi'(0)$.
$\Phi'(0)$ denotes the derivative of the wave function at the origin and $Q=2/3$ the charmed quark electric charge. $p_1$ and $p_2$ are the momenta, $\epsilon^1$ and $\epsilon^2$ the polarization vectors of the photons. $\epsilon$ stands for the polarization vector (tensor) in the case of $\chi_1$ ($\chi_2$). Note, that this result is, strictly speaking, only valid in the short-distance limit, that means for both $p_1^2$ and $p_2^2$ sufficiently different from $M^2$, the squared mass of the bound state.

The amplitude for electron-positron annihilation into the resonances can then be cast into the form
\begin{eqnarray}
\label{eq5}
A(e^+ e^- \to^3P_J)&=&ie^2\int\frac{dp_1}{(2\pi)^4}\bar{v}(l_+)\gamma_{\nu}\not{h}\gamma_{\mu}u(l_-)\frac{1}{h^2}\frac{1}{p_1^2}\frac{1}{p_2^2}A_J^{\mu\nu}(p_1,p_2,\epsilon),
\end{eqnarray}
with $h=l_- - p_1$. Here and in the following the approximation of vanishing electron mass has been made throughout.

Without any further assumption the amplitudes for the electron-positron-$\chi$ coupling can be written in the form
\begin{equation}
A(e^+ e^- \to\kern+4pt^3P_0)=0,
\end{equation}
\begin{equation}
A(e^+ e^- \to\kern+4pt^3P_1)=g_1\bar{v}\gamma_5/{\kern-5pt\epsilon}u,
\end{equation}
\begin{equation}
A(e^+ e^- \to\kern+4pt^3P_2)=g_2\bar{v}\gamma^{\mu}u\epsilon_{\mu\nu}(l_+^{\nu}-l_-^{\nu})/M_{\chi_{2}}.
\end{equation}
In the short distance approximation, using the amplitudes from equations \ref{eq2}, \ref{eq3},
the coefficients characterizing the production rate are given by
\cite{Kuhn:1979bb}
\begin{equation}
g_1=-\frac{\alpha^2\sqrt{2}}{M_{\chi_{1}}^{5/2}}32a\log\frac{2b_1}{M_{\chi_{1}}} \label{eq9}
\end{equation}
\begin{equation}
g_2=\frac{\alpha^2}{M_{\chi_{2}}^{5/2}}64a [\log\frac{2b_2}{M_{\chi_{2}}}+\frac{1}{3}(i\pi+\log{2}-1)]\label{eq10}.
\end{equation}
with the binding energies defined as $b_i=2m-M_{\chi_i}$. The electronic widths which, of course, also characterize the production in $e^+e^-$ annihilation, are given by
\begin{equation}
\Gamma(^3P_1 \rightarrow e^+e^-)=\frac{1}{3}\frac{|g_1|^2}{4\pi}M_{\chi_{1}},
\label{gam1e}
\end{equation}
%(\frac{2}{3})^{2/3}
\begin{equation}
\Gamma(^3P_2 \rightarrow e^+ e^-)=\frac{1}{5}\frac{|g_2|^2}{8\pi}M_{\chi_{2}}.
\label{gam2e}
\end{equation}
and the model dependence is evidently relegated to the couplings $g_1$ and $g_2$. For numerical estimates $|\Phi'(0)|^2=0.1 {\rm GeV}^5$ will be taken.

For the charmed quark mass we take $m_i=\frac{b+M_{\chi_i}}{2}$, the relative size of the absorptive part for negative binding energy (b=-0.5 GeV) is given by 6.26 for $\chi_1$ and 9.34 for $\chi_2$. For positive binding energy b=0.5 GeV the relative size of the absorptive part is given by 0.0 for $\chi_1$ and 0.58 for $\chi_2$. For negative values of $b$ the term proportional to the imaginary part of $\log 2b/M_\chi$ simulates the contribution from the intermediate state $J/\psi + \gamma$, the term proportional $i\pi$ in $g_2$ originates from the two-photon intermediate state.

\begin{table}
\centering
\caption{Electronic widths for $b=-0.5$ GeV and $b=0.5$ GeV.}
\label{npb_tab1}
\begin{tabular}{lll}
\hline
&$\Gamma(\chi_{1}\rightarrow e^+ e^-)$ & $\Gamma(\chi_{2}\rightarrow e^+ e^-)$ \\
\hline
&\multicolumn{2}{l}{$b=0.5$ GeV} \\
\hline
Leading term & 0.0226\ eV & 0.0243\ eV \\
exact result &0.0317\ eV & 0.0159\ eV\\
\hline
&\multicolumn{2}{l}{$b=-0.5$ GeV}\\
\hline
Leading term &0.164\ eV & 0.0512\ eV \\
exact result &0.141\ eV & 0.0731\ eV\\
\hline
\end{tabular}
%\vspace*{1cm}
\end{table}
In view of the strong dependence of the rate on subleading terms one may try to take corrections resulting from the the binding energy into account.
This implies the inclusion of terms of order $1-x$ with $x=4m^2/M_{\chi_i}^2$ and one finds for the couplings
\begin{equation}
g_{1_{\gamma \gamma}}=  \frac{16\alpha^2a}{\sqrt{m}M_{\chi_{1}}^2}\Bigg[ 
  \log\left(\frac{x}{1+x}\right)
   \left(1-x\right)
  -\left(\log\left(\frac{x}{1-x}\right)+i\pi\right)
   \left(1+x\right)\Bigg] ,
\label{eq13}
\end{equation}
\begin{eqnarray}
  g_{2_{\gamma \gamma}}&=&   \frac{32\sqrt{2}\alpha^2a}{3\sqrt{m}M_{\chi_{2}}^2} 
 \Bigg[\left( \frac{1+x}{2}+\frac{8}{(1+x)^2}\right)\log(1-x)+ \frac{3}{2}\left(1+x\right) \log(1+x)\nonumber\\
 &&-2\left(1+x + \frac{2}{(1+x)^2}\right)\log(x)-\frac{8}{(1+x)^2}\log(2)-1 
  -\frac{i\pi}{2}\left( 1+x +\frac{8}{(1+x)^2}\right)
\Bigg]\nonumber.\\
\label{eq14}
\end{eqnarray}
which, in the limit $x\to 1$ evidently reduces to the values given in eqs. \ref{eq9} and \ref{eq10}. The results for negative and positive $b$ and both $x\approx 1$ and $x\neq 1$ are listed in Table \ref{npb_tab1}.

Up to this point the short-distance approximation has been taken throughout. However, considering the fact that binding energies and overall mass scale are comparable, different assumptions on the form factor were made in the literature. In \cite{Czyz:2016xvc} an ansatz for the form factor has been made which in the case of $\chi_1$ takes the $\chi_1\psi\gamma$ and $\chi_1\psi'\gamma$ couplings into account, in the case of $\chi_2$ in addition the coupling to the $\gamma\gamma$ intermediate state.
\begin{figure}[h]
% Use the relevant command for your figure-insertion program
% to insert the figure file.
\centering
\includegraphics[width=8cm,clip]{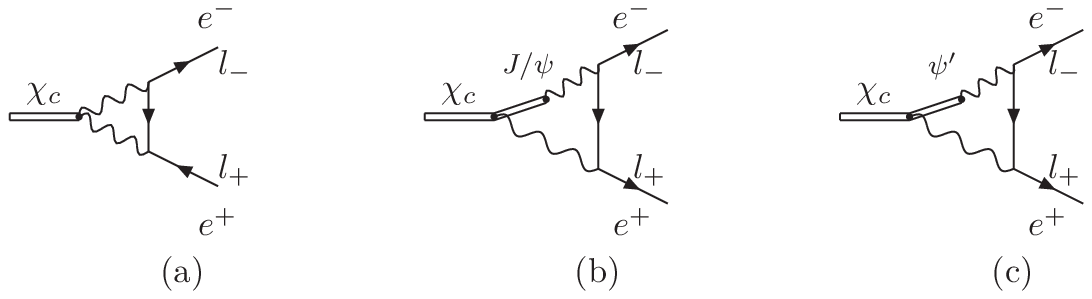}
\caption{Diagrams for decay widths $\Gamma(\chi_{{0,1,2}\to e^+ e^-})$.}
\label{fig-2}       % Give a unique label
\end{figure}
The $R$ value at the peak of the cross section is given by
\begin{equation}
R_{peak}=\frac{\sigma_{res}^{(0)}}{\sigma_{pt}}=\frac{\Gamma_{ee}}{\Delta}\frac{9}{4\alpha^2}\sqrt{2\pi}\frac{\Gamma_{had}}{\Gamma_{tot}}N_Z.
\end{equation}
In this formulation interference contributions with the continuum are neglected. The symbols $\Gamma_e$, $\Gamma_{\rm had}$ and
$\Gamma_{\rm tot}$ denote the width of the resonance into $e^+e^-$, into hadrons and the total width.
(Note that $\Gamma_{had} = 1 - \Gamma_{J/\psi\gamma}$.)
$\Delta\approx 2$ MeV stands for the machine energy resolution.
Taking for illustration values of $\Gamma_e$ between 0.1~eV and 0.5~eV one finds an enhancement of the $R$-value between $2 \cdot 10^{-3}$ and
$1 \cdot 10^{-2}$.

Note that the $1^{++}$ state receives a (small) additional contribution from the axial vector part of the neutral current \cite{Kaplan:1978wu,Yang:2012gk,Czyz:2016xvc},
which also contributes to the reaction $e^+e^-\to \chi_1$.
To identify the interference, term the neutral current amplitude has to be decomposed into the form
$(V_e + A_e) A_c$. It is then the interference between the $A_e A_c$ term from the neutral current and the real part of the
electromagnetic amplitude which contributes to the rate. Specifically
\begin{eqnarray}
\Gamma(\chi_{1} \rightarrow e^+e^-)&=&\frac{M_{\chi_{1}}}{3\pi}\Bigg[\frac{|g_1|^2}{4}+\frac{aG_F}{\sqrt{2m}Q^2} \textrm{Re}(g_1)+\frac{a^2G_F^2}{mQ^4}\Bigg(1-4\sin^2{\theta_W}+8\sin^4{\theta_W}\Bigg) \Bigg],
\label{nZcon}
\end{eqnarray}
where $G_F$ is the Fermi constant and $\theta_W$ the weak mixing angle. The coupling $g_1$ has been defined in eqs. \ref{eq13}.

In this extended model the electronic widths were calculated using the diagrams shown in Fig. \ref{fig-2}, plus the neutral current piece given above.
The couplings $g_1$ and $g_2$ are obtained by performing the loop integrals and can be divided into three parts
\begin{equation}
g_i=g_{i_{\gamma\gamma}} + g_{i_{J/\psi \gamma}} + g_{i_{\psi^{\prime}\gamma}}
\end{equation}
with contributions coming from Figs. \ref{fig-2}a, b and c. The results for this model are listed in Table \ref{tab2}. Note that in this model the $\chi_2$ state gives a particularly large value for the electronic width. 
\begin{table}
\centering
\caption{Electronic widths for $\chi_{1}$ and $\chi_{2}$. See text for details.}
\label{tab2}
\begin{tabular}{cccccc}
\hline
 & $\gamma \gamma+J/\psi\gamma+\psi'$ & $\gamma \gamma$ & $J/\psi\gamma$& $\psi'$ & QED+$Z^0$ \\ 
\hline
$\Gamma_{(\chi_{1}\rightarrow e^+ e^-)}$ [eV]&0.42  &0.102 &0.007& 0.094 & 0.41 \\
$\Gamma_{(\chi_{2}\rightarrow e^+ e^-)}$ [eV]&4.25  &0.004  & 1.41& 0.448 &-\\
\hline
\end{tabular}
\end{table}

\section{The process $e^+e^- \to \chi_i \to \gamma J/\psi (\to \mu^+\mu^-) $}
As mentioned already in the Introduction, the search for direct $\chi_J$ production may either be based on its hadronic decay and the corresponding search of a resonance enhancement of the $e^+e^-$ cross section at
$\sqrt{s} = M_{\chi_i}$. Alternatively one may search for a resonance enhancement at this energy in the reaction
$e^+e^-\to \chi_i \to \gamma J/\psi (\to \mu^+\mu^-)$. In view of the relatively large branching ratios
${\rm Br}(\chi_1 \to \gamma J\psi)= 33.9 \pm 1.2\%$ and
${\rm Br}(\chi_2 \to \gamma J\psi)= 19.2 \pm 0.7\%$
and the clean signal a slight enhancement might be visible for the final state $\gamma J/\psi$, when $\sqrt{s}$ is varied in the vicinity of
$M_{\chi_J}$.

\begin{figure}[h]
% Use the relevant command for your figure-insertion program
% to insert the figure file.
\centering
\includegraphics[width=8cm,clip]{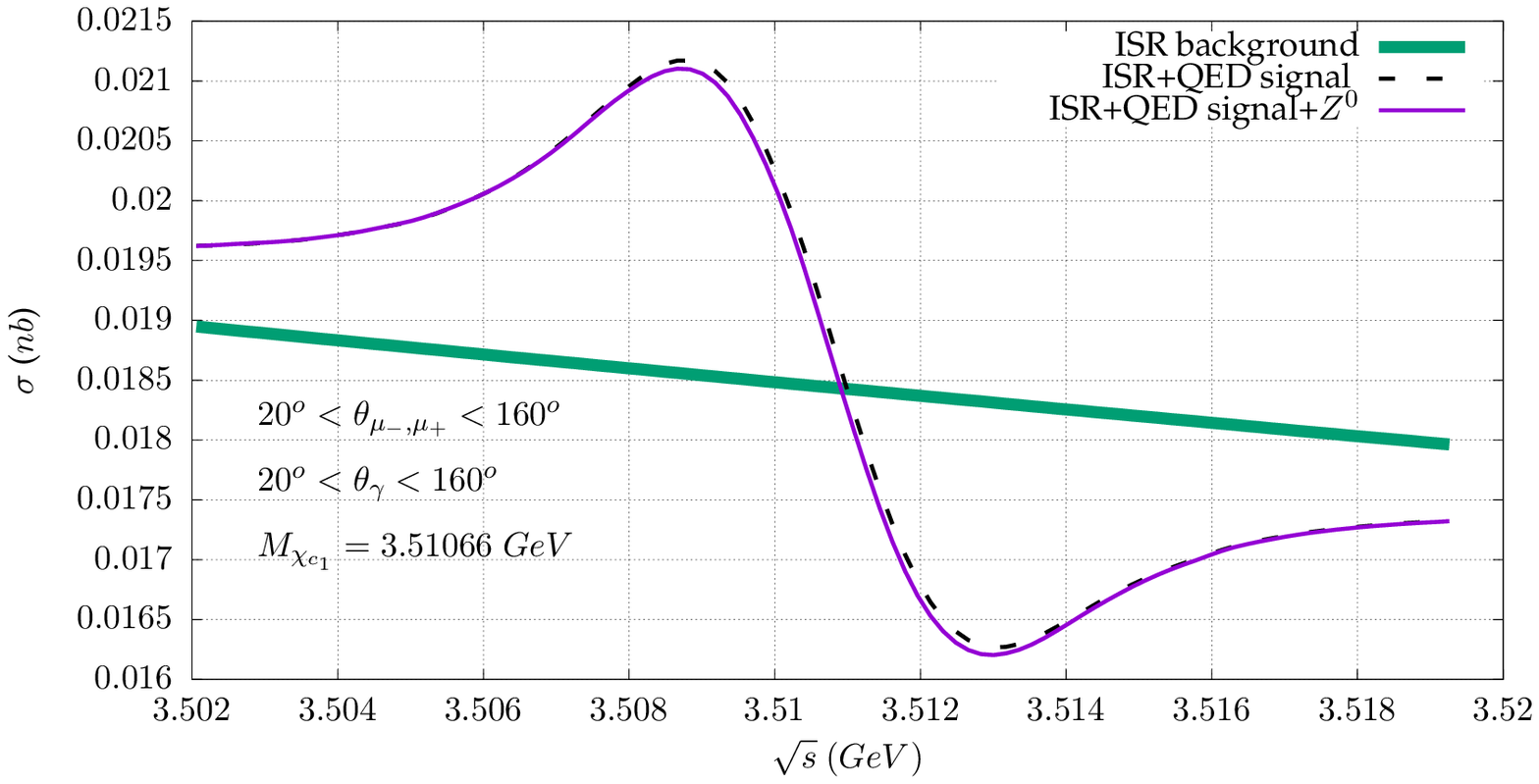}
\caption{The cross section $e^+ e^- \to \mu^+ \mu^- \gamma$ in the $\chi_1$ - region.}
\label{fig-3}       % Give a unique label
\end{figure}

\begin{figure}[h]
% Use the relevant command for your figure-insertion program
% to insert the figure file.
\centering
\includegraphics[width=8cm,clip]{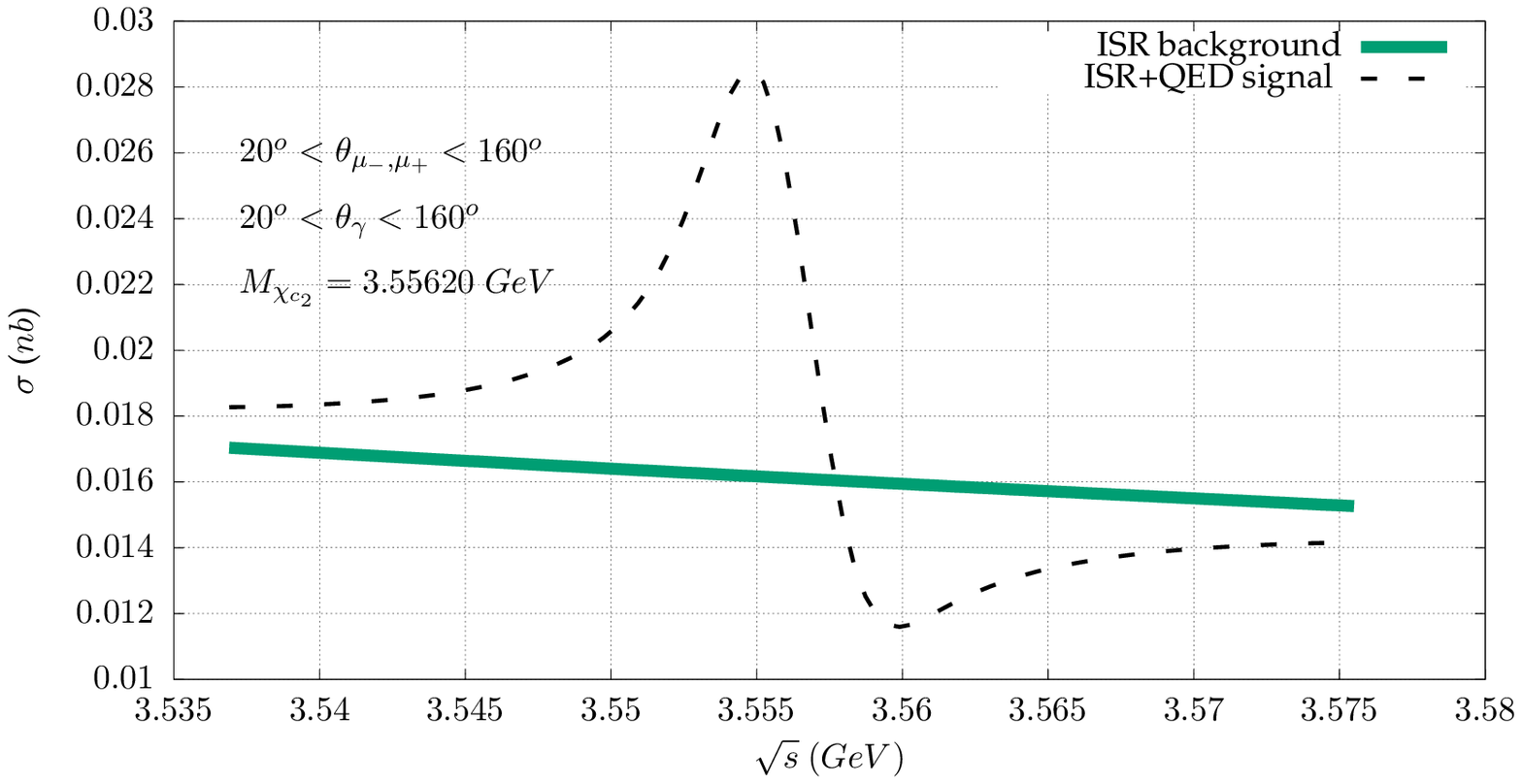}
\caption{The cross section $e^+ e^- \to \mu^+ \mu^- \gamma$ in the $\chi_2$ - region.}
\label{fig-4}       % Give a unique label
\end{figure}

\begin{figure}[h]
% Use the relevant command for your figure-insertion program
% to insert the figure file.
\centering
\includegraphics[width=8cm,clip]{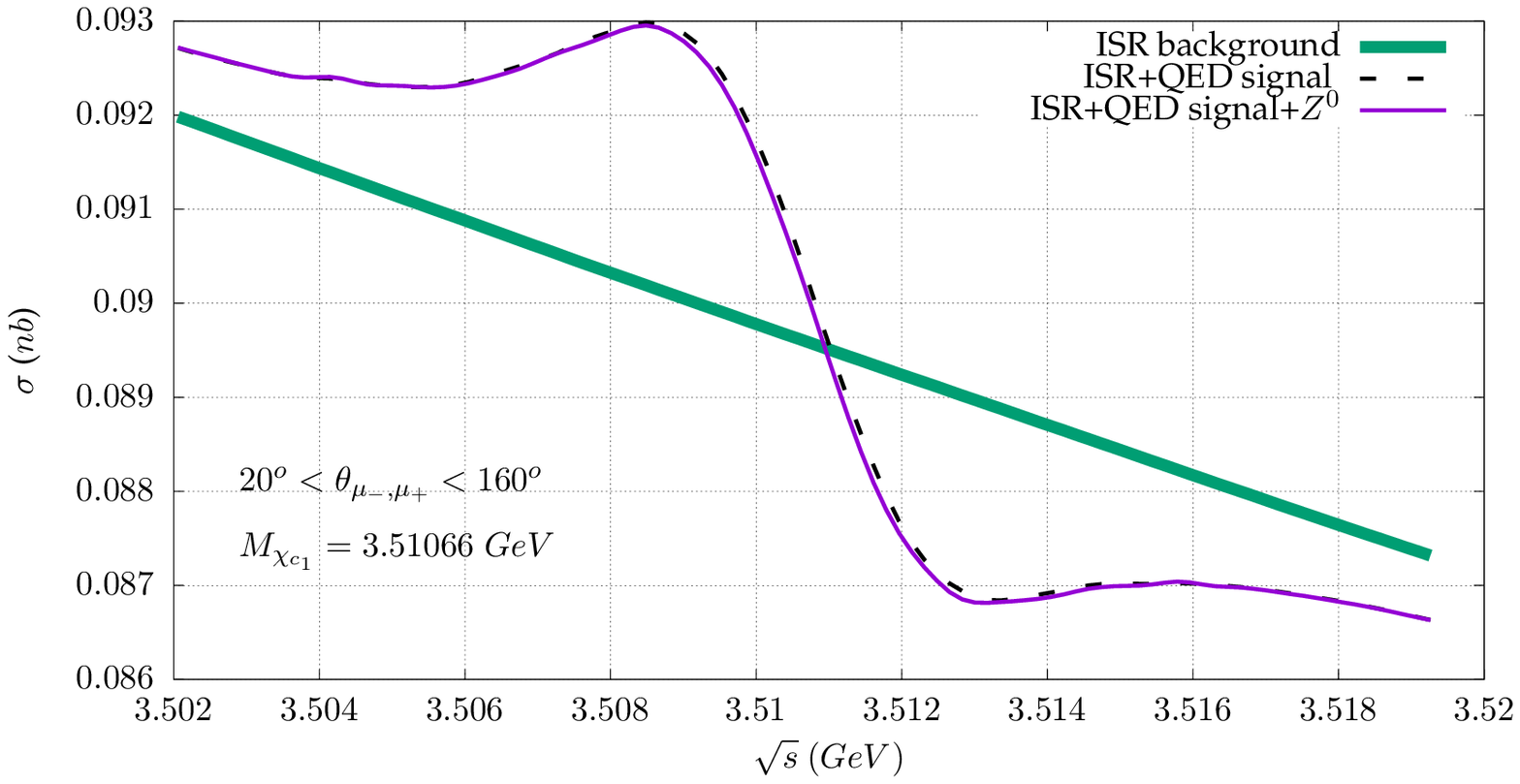}
\caption{The cross section $e^+ e^- \to \mu^+ \mu^- \gamma$ in the $\chi_1$ - region.}
\label{fig-5}       % Give a unique label
\end{figure}

\begin{figure}[h]
% Use the relevant command for your figure-insertion program
% to insert the figure file.
\centering
\includegraphics[width=8cm,clip]{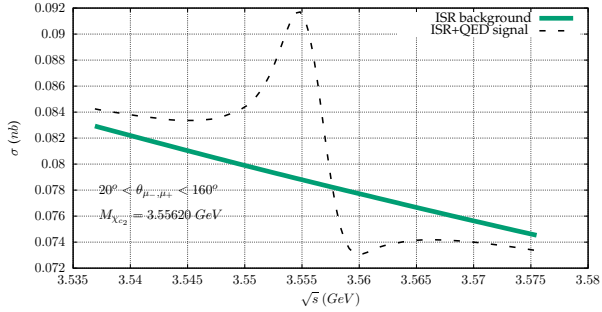}
\caption{The cross section $e^+ e^- \to \mu^+ \mu^- \gamma$ in the $\chi_2$ - region.}
\label{fig-6}       % Give a unique label
\end{figure}

Predictions for the combined cross section of the reaction
$e^+e^- \to \mu^+\mu^- \gamma$ in the vicinity of the $\chi_1$ and $\chi_2$ resonances, respectively, are shown in Figs. \ref{fig-3}, \ref{fig-4}, \ref{fig-5} and \ref{fig-6}.
Figs. \ref{fig-3} and \ref{fig-4} give predictions for the cross section in the $\chi_1$ and $\chi_2$ regions, respectively, including cuts on photon and lepton angles
inside the detector region, Figs. \ref{fig-5} and \ref{fig-6} give the corresponding cross sections without cuts on photon emission. The predicted values would be clearly sufficient for observation at the BESIII experiment, once a scan with energies around the $\chi_1$ and $\chi_2$ resonances would be performed. In these cases a beam energy spread of $1$ MeV with Gaussian distribution was assumed. Contributions from the diagrams depicted in
Figs. 1b and 1a with $J/\psi$ substituted by $\psi'$ are completely negligible since the muon pair mass was chosen to be in the interval
$[M_{J/\psi} - 3 \Gamma_{J/\psi} , M_{J/\psi} + 3 \Gamma_{J/\psi}]$.

After these cuts a signal of up to $75\%$ for $\chi_2$ and up to $14\%$ for $\chi_1$ compared to the radiative return
background could be observed. In fact the BESIII collaboration should be able to measure these cross sections and extract the electronic widths of $\chi_1$ and $\chi_2$, if the optimistic assumptions will turn out to be correct. In this case the scan in the vicinity of the resonances would provide the possibility to test the model and, furthermore, to extract the phase between radiative return and direct production of the resonances.

\newpage
\section{Conclusions}
Direct resonant production of of $\chi_1$ and $\chi_2$ states could lead to a measurable enhancement of the cross section in electron-positron annihilation at the BESIII storage ring. The predictions exhibit a sizable model dependence and can be considered to be of qualitative nature only. Nevertheless, a resonant signal both in the hadronic cross section and in the $\mu^+\mu^-\gamma$ channel could be observed under favorable circumstances.

\end{document}